\documentclass[journal]{IEEEtran}
\usepackage[pdftex]{graphicx} 
\graphicspath{ {../png/} }
\pdfoutput=1
\usepackage[utf8]{inputenc}
\usepackage{csquotes}
\usepackage[english]{babel}
\usepackage[style=ieee, sorting=none]{biblatex}
\usepackage{amsmath}
\usepackage{amssymb}
\usepackage{authblk}
\usepackage{float}
\usepackage{bm}
\usepackage{bookmark}

\begin{document}

\title{A CT Image Denoising Method Based on Projection Domain Feature}
\author{Mengyu Sun, Dimeng Xia, Shusen Zhao, Weibin Zhang and Yaobin He
  \thanks{\textit{(Corresponding authors: Dimeng Xia; Yaobin He.)}}
  \thanks{Mengyu Sun, Dimeng Xia, Shusen Zhao and Weibin Zhang are with National Center for Applied Mathematics Shenzhen (NCAMS), Southern University of Science and Technology, Shenzhen 518055, China (e-mail: sunmy@mail.sustech.edu.cn; xiadm@sustech.edu.cn; zhaoss@sustech.edu.cn; zhangwb3@sustech.edu.cn).}
  \thanks{Yaobin He is with CETC Key Laboratory of Smart City Modeling Simulation and Intelligent Technology, The Smart City Research Institute of CETC, Shenzhen 518055, China (e-mail: ho.frank@qq.com).}}

\maketitle

\begin{abstract}
In order to improve image quality of projection in industrial applications, generally, a standard method is to increase the current or exposure time, which might cause overexposure of detector units in areas of thin objects or backgrounds. Increasing the projection sampling is a better method to address the issue, but it also leads to significant noise in the reconstructed image. This paper proposed a projection domain denoising algorithm based on the features of the projection domain for this case. This algorithm utilized the similarity of projections of neighboring veiws to reduce image noise quickly and effectively. The availability of the algorithm proposed in this work has been conducted by numerical simulation and practical data experiments.
\end{abstract}


\section{Introduction}
\IEEEPARstart{C}{omputer} Tomography (CT) imaging has been widely used in industrial non-destructive testing and clinical examination due to its
 intense penetration. However, in industrial CT testing, especially for high-attenuation materials (such as rock core, metal casting, etc.) or the length-to-width ratio of the object measured is highly different (such as power battery, circuit board, etc.), to improve the quality of the reconstructed image and to ensure that the detector still obtains sufficient photons at sampling views which are difficult to get through, it is expected to increase the exposure time for a single projection or to increase the ray tube current, etc. In practice, there is an upper limit to the number of counts obtained from the detector unit, so when the X-ray passes through the low-attenuation object or no-object area, a long integration time or a large tube current usually leads to the count on the detector exceeding its maximum, resulting in “overexposure”, which is difficult to obtain the ideal reconstructed image. To solve the above problem, common methods are used to reduce the exposure time of one projection while increasing the number of sampling views, ensuring the data is available. The noise level of the obtained projection data is relatively high. Using traditional CT reconstruction algorithms (ART, FBP, etc.) to reconstruct this kind of projection data directly, the CT image's quality is low, affecting the subsequent qualitative and quantitative image analysis.

When the projection data is not ideal, researchers have proposed a variety of methods to suppress noise in the reconstructed image. One kind of method establishes an optimization model with a data fidelity term and a regularization term to suppress noises by solving the model. Based on the theory of compressive sensing (CS) \cite{2006Donoho}, several regularization methods have been proposed for low-dose CT reconstruction, for example, regularization based on discrete gradient transformation denoising \cite{2009Accurate}, non-local wavelet transform denoising \cite{2009Wavelet}, \cite{2013Adaptive}, \cite{2022Asymptomatic}, dictionary learning-based denoising. Total variation (TV) minimization denoising \cite{2014Non}, \cite{2021Mutual} aims to minimize the total variation of the image to the maximum extent while preserving its features, which outputs high-quality denoising results with proper adjustment of the regularization parameters, that is, reducing noises with preserving edges and clear features. Non-local filtering, such as non-local mean, block-matching, and 3D filtering (BM3D), utilizes the highly redundant information typically present in images, which compares the similarity of patches across the image and assigns the weighted average value of adjacent pixels to a certain pixel to reduce noises \cite{2011Sparse}, \cite{2014Adaptive}, \cite{2010Block}, \cite{2017Applications}, \cite{2018Improved}, \cite{2019Ultra}. Dictionary learning-based denoising \cite{2017Z}is a machine learning‐based approach that formats a dictionary of image patches based on a batch of training images and then uses this dictionary to denoise new images. However, there are some limitations in terms of large computation and complex optimization adjustments that cause little effect or over-smoothness and detriment to retain rich texture information of tissue structure. Besides, robustness and generalization of the above traditional denoising methods also need to be improved.

In recent years, deep learning (DL) techniques have become one of the research hotspots and have shown great potential in image processing tasks. Multiple DL approaches have been proposed for CT image denoising, including but not limited to Convolutional Neural Networks (CNN), Generative Adversarial Networks (GANs), Variational Autoencoders (VAEs), Deep Residual Networks (ResNets), Transformer-based methods, Attention-based Networks, and hybrid networks combined with various methods, among which CNN and GANs are considered as the predominant CT denoising models \cite{2024CT}. DL approaches have outperformed traditional denoising methods but still shown limitations: one of the primary challenges is the strong dependence on reference datasets, which consist of large amounts of high-quality and consistent prior data \cite{2018Deep}, \cite{2021A}, \cite{2020SACNN}, \cite{2022Performance}, \cite{2018Sharpness}. In addition, DL methods mostly focus on denoising in the image domain, neglecting the structural features of the object in the projection data.

The typical projection domain denoising method is sinogram domain filtration, which includes sinogram filtering \cite{2013Adaptive}, bilateral filtering \cite{2022Asymptomatic}, and penalty-weighted least squares \cite{2014Non}. These kinds of approaches are based on direct smoothing of the sinograms, resulting in a loss of spatial resolution \cite{2012Ray}. Iterative reconstruction (IR) performs multiple iterations for the optimization of the objective function, that is, alternating the operations of forward and backward projections between the image domain and projection domain until the objective function minimizes convergence and ultimately outputs low-noise reconstruction images \cite{2012Noise}. More iterations mean longer processing time, especially for mixed iterations; thus, IR usually requires massive computation and sufficient time. To better utilize projection domain features, some approaches have been proposed that estimate the types of noises in the projection data, for example, simulating the extracted noises as Poisson noise and non-stationary Gaussian noise, then construct corresponding optimization models to achieve noise reduction in the projection domain \cite{2019Projection}. This kind of method generally needs to estimate the parameters of the established model, which may not obtain the ideal image when the parameters are inaccurate.

With the intent of solving the above problem, we propose a CT image denoising algorithm based on projection domain features. Based on the similarity of projection data collected at neighboring views, this algorithm utilizes the singular value decomposition (SVD) method to preserve the maximum similarity feature and exclude noise information to suppress noise from projection data. This method is free of prior information, making it preferable practicality, and can be embedded into other denoising algorithms flexibly to obtain higher quality CT reconstruction images.

\section{Methods}

\subsection{Feature Extraction}
The feature extraction involves two major steps: grouping the projections and performing SVD. With adequate sampling quantity, the projections among several neighboring views show high similarity. Thus, the projections can be grouped and divided into several submatrices based on views. Due to the similarity of neighboring projections, each submatrix can be approximated as a matrix with the rank 1. Then, the feature matrix is formed after performing SVD on each submatrix.

Assume that \(\bm {P}\in{R^{M\times N}}\)  denotes the CT projection, where $M$ is the number of detector elements, and $N$ is the number of sampling angles, then the projection matrix $\bm {P}$ can be represented as

\begin{equation}
\bm {P}={(\bm {p}_1,\bm {p}_2,\bm {p}_3,\cdots,\bm {p}_n,\cdots,\bm {p}_N)}^T
\end{equation}
where \(\bm {p}_n=(u_1,u_2,u_3,\cdots ,u_m,\cdots ,u_M)\) denotes the projection vector under each sampling view. Based on the similarity of projections mentioned above, we can redefine the projection matrix $\bm {P}$ according to each group contains \(t\in{Z^+}\) sampling views, Eq. (1) is rewritten as

\begin{equation}
\bm {P}={(\bm {q}_1,\bm {q}_2,\bm {q}_3,\cdots,\bm {q}_s,\cdots,\bm {q}_{N/t})}^T
\end{equation}
where \(\bm {q}_s\in{R^{M\times t}}\). The mathematical expression of SVD of $\bm {q}_s$ is

\begin{equation}
\bm {q}_s=\bm {UQV}^T
\end{equation}
where \(\bm {q}_s\in R^{m\times n}\), \(\bm {Q}\in R^{m\times n}\) is a low-rank diagonal matrix, \( \bm {U}\in R^{m\times m}\) and \(\bm {V}\in R^{n\times n}\)  are the left singular matrix and the right singular matrix respectively. The elements on the main diagonal in $\bm {Q}$ are arranged descending order along the diagonal direction and are considered as the eigenvalues of $\bm {q}_s$. The key feature of $\bm {q}_s$ represents the most important feature of $\bm {Q}$.

\subsection{Noise Removal}
The noise removal involves two major steps: screening eigenvalues and restoring the projection matrix. Considering the sources of information in projection, the key feature mentioned above of $\bm {q}_s$ is considered as structure information of the detected object, while noise is an unimportant feature. Thus, eigenvalue filtering contributes to noise suppression. The strategy of eigenvalue screening is determined based on the value of the eigenvalu e and the grouping of the projection matrix, taking the former as an example, the eigenvalue screening is expressed by 

\begin{equation}
diag\left( {\bf{Q}} \right) = \left\{ \begin{array}{l}
{\sigma _r},{\rm{ }}{\sigma _r}{\rm{  > }}\varepsilon \\
0,{\rm{   others}}
\end{array} \right.
\end{equation}
where $\varepsilon$ and $\sigma_r$ denote the set threshold parameter and the eigenvalues of $\bm {Q}$, respectively. Once $\varepsilon$ is given, important features preservation and unimportant features elimination are implemented. After screening, the feature matrices $\bm {Q}$ and their corresponding left singular matrices $\bm {U}$, right singular matrices $\bm {V}$ are calculated according to Eq.(3) to generate new submatrices, then the complete and low noise projection matrix is restored. 

\section{Experiments and Results}
In this article, we use numerical simulation and practical experiments and compared the proposed algorithm with existing methods, including the TV algorithm and the noise-level weighted TV regularization algorithm (NMTV) \cite{2019Projection}. All images are reconstructed using the simultaneous algebraic reconstruction technique (SART) algorithm. TV algorithm as a post-processing method is close to practical denoising applications for high-noise images, showing the capability of noise reduction in the image domain. NWTV allows adaptive adjustment of regularization terms with the type and level of spatially varying noise and provides the balance between noise reduction and detail preservation, which is used for comparison within the projection domain processing method.

\subsection{Numerical Simulations}

\begin{figure}
\includegraphics[width=1\columnwidth]{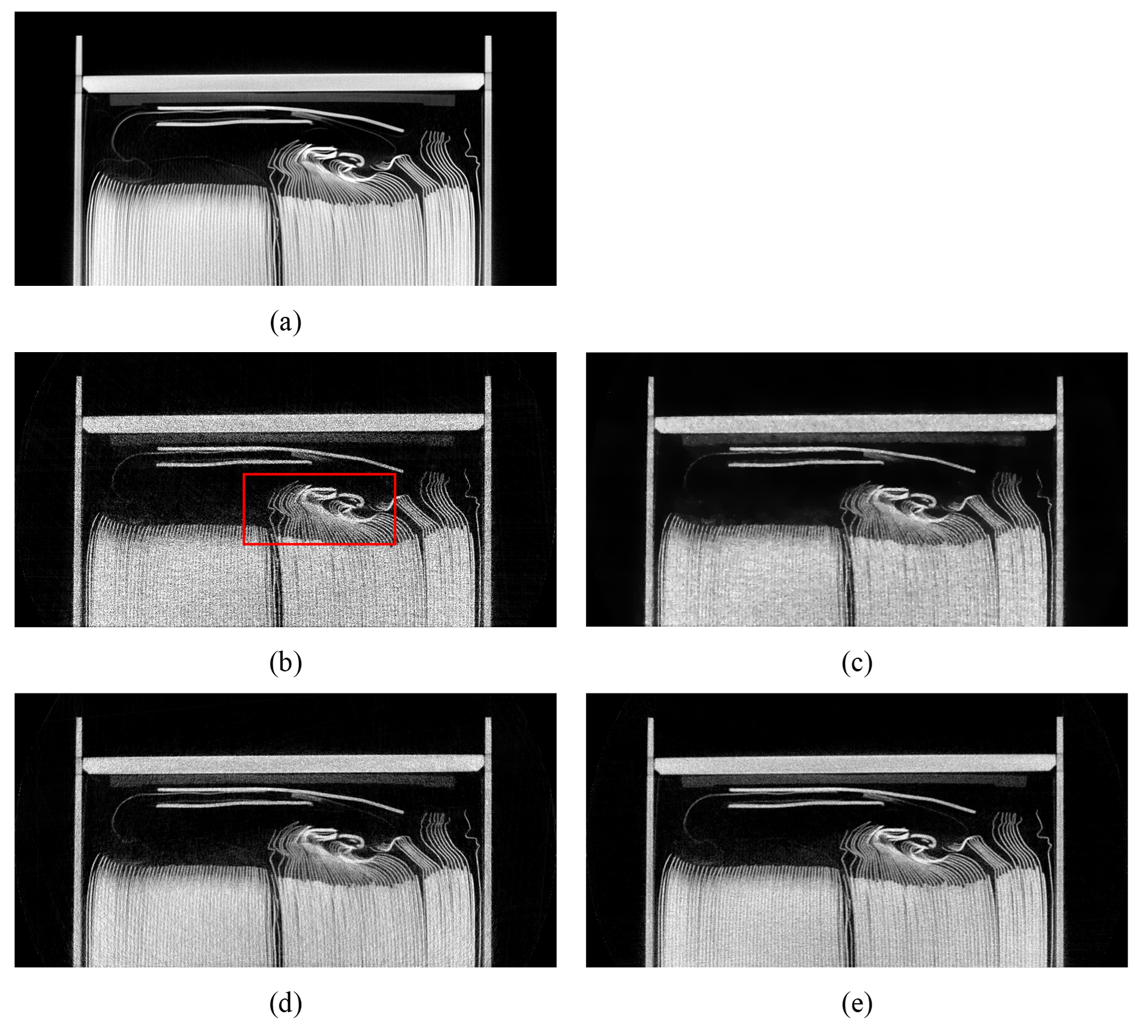}
\centering
\caption{Comparison of denoising results for simulated square shell battery data. (a) reference image, (b) noisy simulation, (c) denoised with TV-L1, (d) denoised with NWTV, (e) denoised with the proposed algorithm. All images are reconstructed by SART and displayed with window level [0, 1.1].}
\label{fig1}
\end{figure}

To verify the effectiveness of the proposed method in industrial CT application, a simulated dataset of square shell battery (35mm thick) is used, its reference image as shown in Figure ~\ref{fig1}(a). The settings of the simulated parameters are as follows: the distance from the X-ray source to the object (SOD) is 400mm, the distance from the X-ray source to the detector (SDD) is 600mm; the pixel matrix of detector is 1024$\times $1024 with 0.1mm pixel size; the 104 initial photons of Poisson noise are embedded in the simulated data; total 1080 frames are collected during 360$^{\circ}$ scanning with 5 inter frame merge. The threshold parameter $\varepsilon$ of the proposed method is set to 50.

\begin{figure}
\includegraphics[width=1\columnwidth]{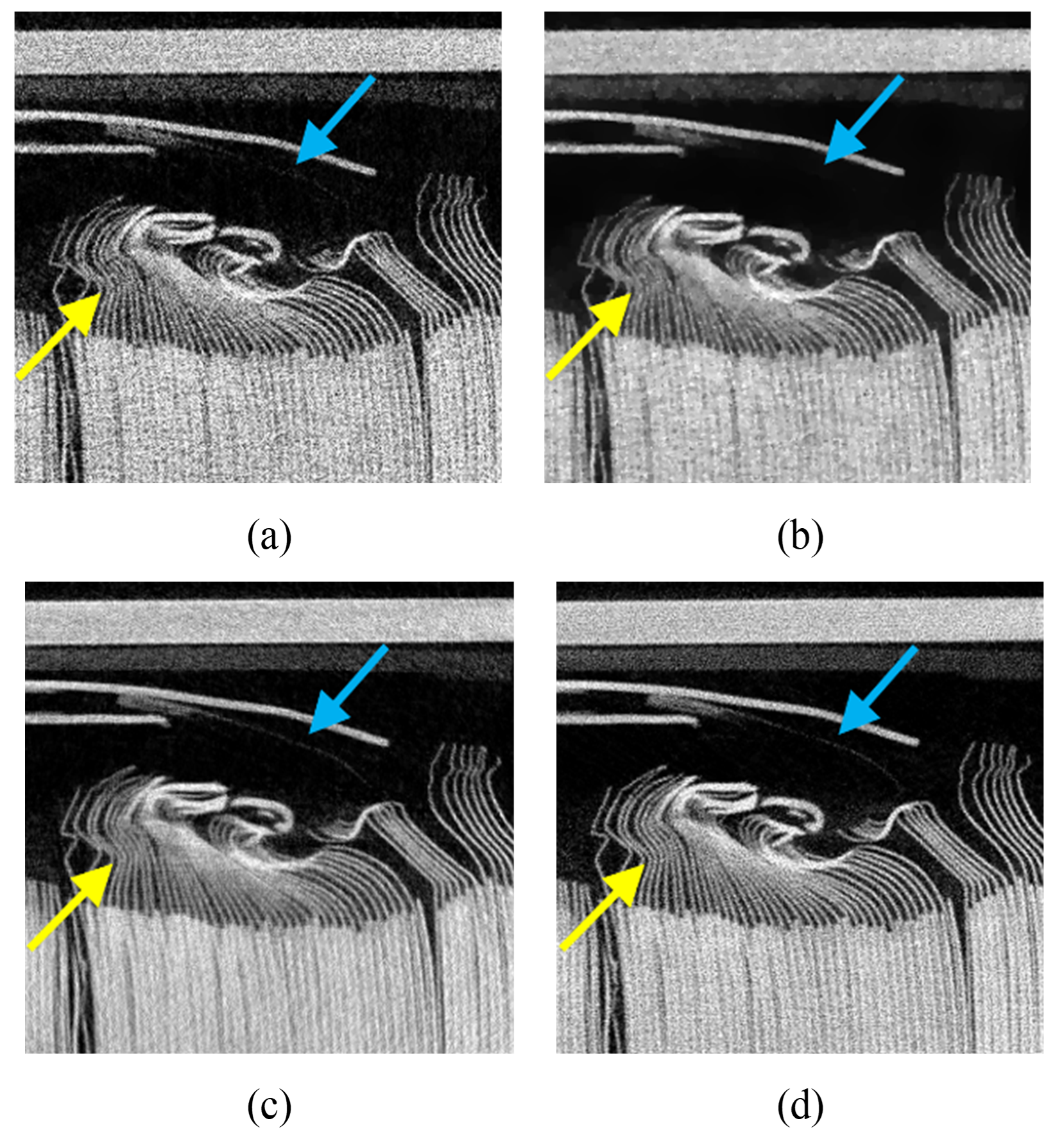}
\centering
\caption{ Partial magnification of each method in Figure 1. (a) noisy simulation, (b) denoised with TV-L1, (c) denoised with NWTV, (d) denoised with the proposed algorithm. The yellow and blue arrows indicate the edges of the structure. All images are reconstructed by SART and displayed with window level [0, 1.1].}
\label{fig2}
\end{figure}

Fig.~\ref{fig1} shows visual denoising effects with different denoising methods in the simulated battery data. Compared to the SART reconstruction image, TV-L1, NWTV and the proposed algorithm perform better to a certain extent in improving noise suppression. We used a red local area box to show the same location area with more details and textures in each image and zoomed in to Fig.~\ref{fig2}. Observing the details of this area, we can intuitively compare the ability of each method to suppress noise and preserve the edges.

As shown in Fig.~\ref{fig2}(b), TV-L1 blurs a multi-layer bending structure marked by the yellow arrow in the lower left corner and loses details of mylar film marked by the blue arrow in the upper right corner. Although NWTV performs similarly to our method, it smooths out the textures marked by arrows in Fig.~\ref{fig2}(c), whereas our algorithm preserved the edges of these structures in Fig.~\ref{fig2}(d) very well.

\subsection{Practical Experiment}

\begin{figure}[htbp]
\centering
\includegraphics[width=0.7\columnwidth]{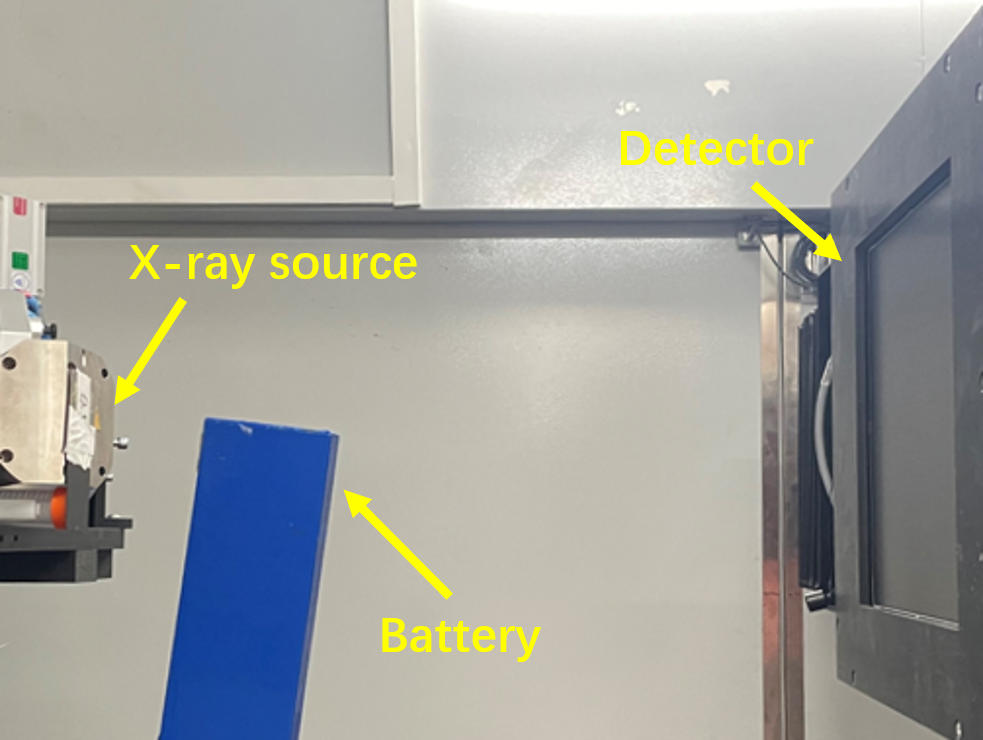}
\caption{Experimental setup of the battery scanning.}
\label{fig3}
\end{figure}

For practical experiments, we employed the scanning dataset of power battery (width$\times $length$\times $height, 104$\times $305$\times $33mm), the battery is imaged on an industrial CT system (Fig.~\ref{fig3}).The settings of the simulated parameters are as follows: the voltage is 220kV and the current is 0.4mA; the exposure time is 0.25s; SOD is 230mm, SDD is 660mm; the pixel matrix of detector is 3072$\times $3072 with 0.1mm pixel length; and total 4500 projections are collected during whole circle scanning.

\begin{figure}
\includegraphics[width=1\columnwidth]{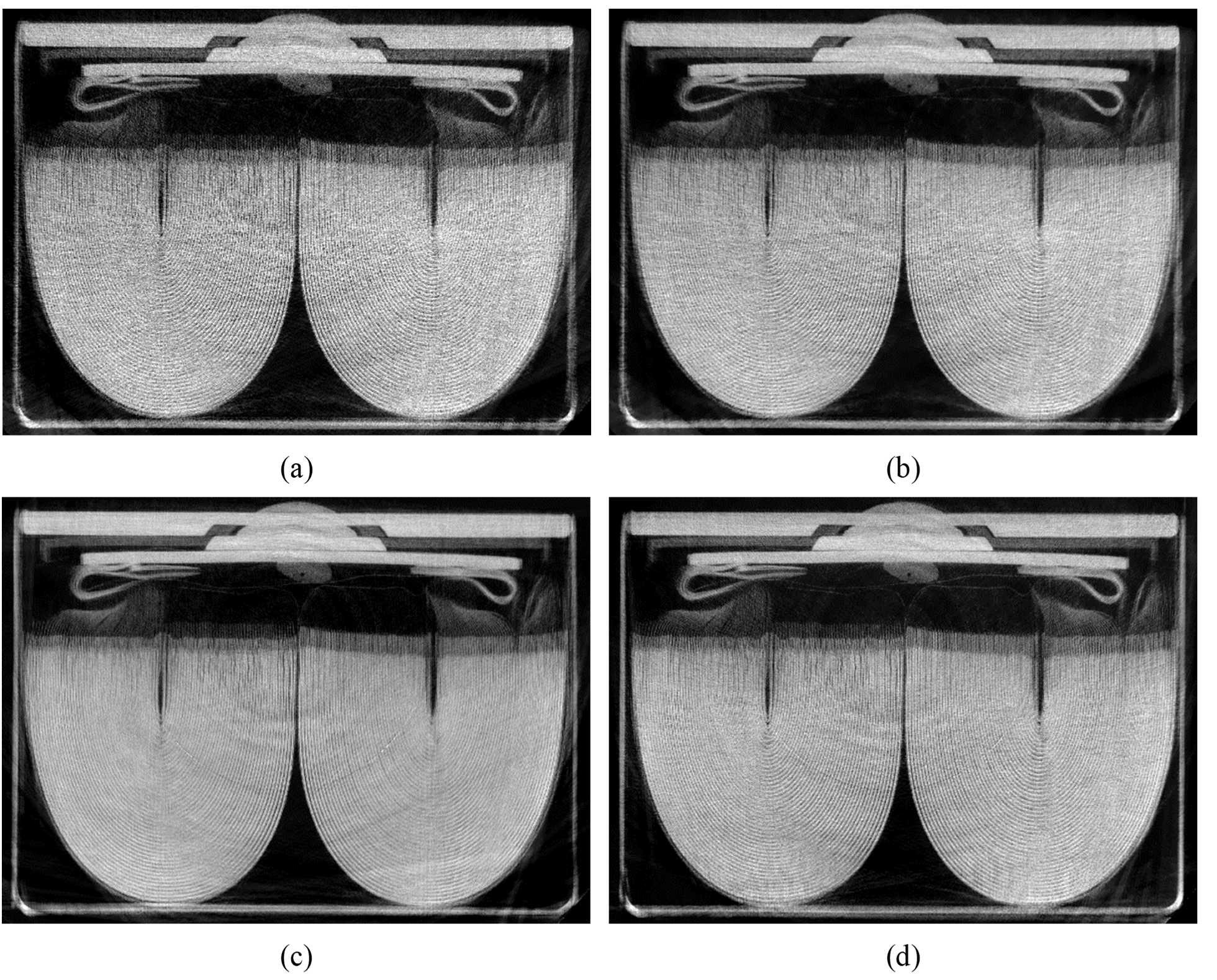}
\centering
\caption{Comparison of denoising results for collected power battery data. (a) reconstruction result without process, (b) denoised with TV-L1, (c) denoised with NWTV, (d) denoised with the proposed algorithm. All images are reconstructed by SART and displayed with window level [0, 0.05].}
\label{fig4}
\end{figure}

\begin{figure}
\includegraphics[width=1\columnwidth]{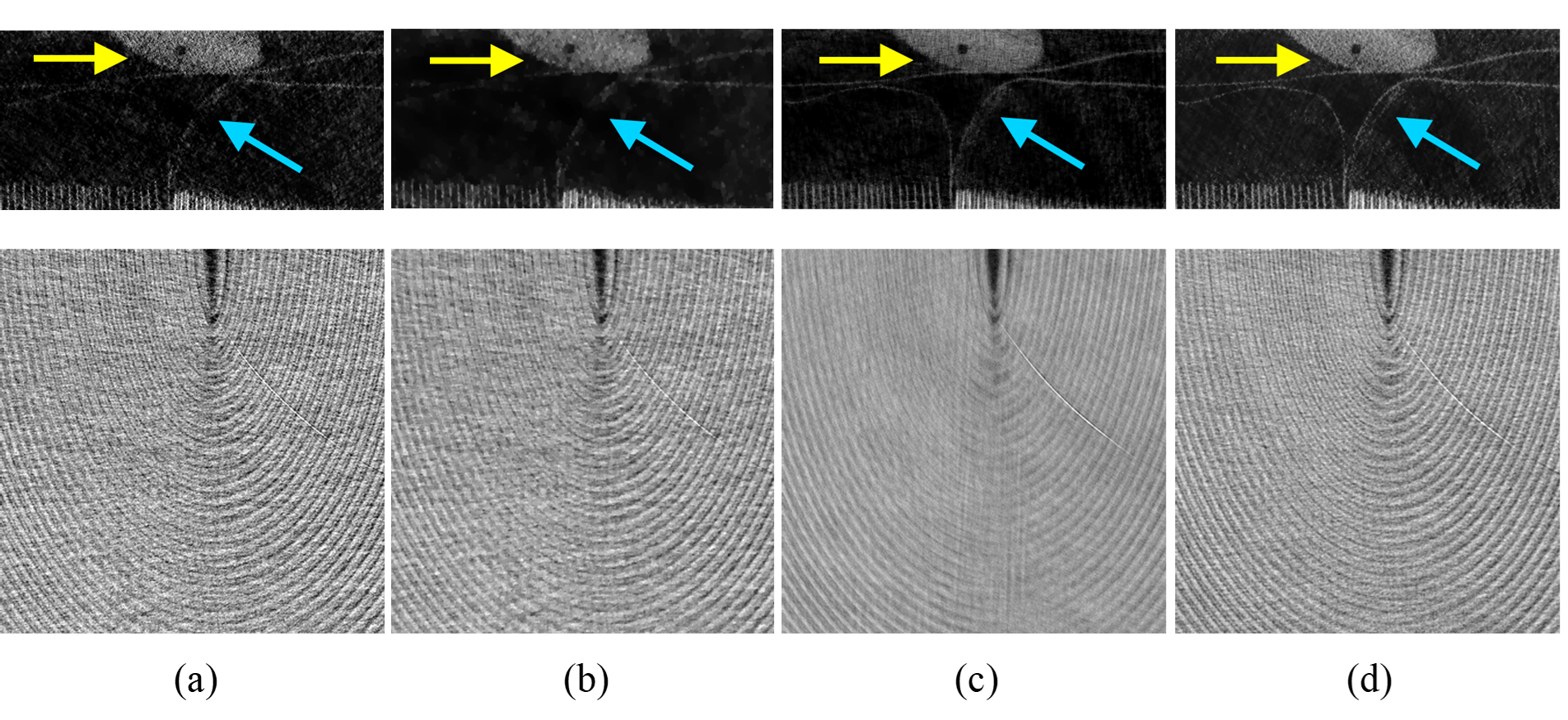}
\centering
\caption{Partial magnification of each method in Figure 4. (a) reconstruction result without process, (b) denoised with TV-L1, (c) denoised with NWTV, (d) denoised with the proposed algorithm. The yellow and blue arrows indicate the edges of the structure. All images are reconstructed by SART and displayed with window level [0, 0.04]. }
\label{fig5}
\end{figure}

Fig.~\ref{fig4} and Fig.~\ref{fig5} report the comparison of reconstruction results from different methods. Severe noise in the reconstruction result without processing result in details marked by the blue and yellow arrows being obscured by noise, as shown in Fig.~\ref{fig4}(a) and Fig.~\ref{fig5}(a). In this case, TV-L1 and NWTV present noticeable noise reduction, but poor detail preservation especially the over-smoothness of stacked areas in the lower panel of Fig.~\ref{fig5}(c), which is similar with simulation results. In contrast, our proposed algorithm captures fine edges better, while effectively suppressing the noise.

\section{Conclusion}
In this paper, we propose a CT image denoising algorithm based on SVD in the projection domain. The algorithm fully utilizes the similarity of projections collected at adjacent angles, performs SVD to retain the maximum similarity feature, and eliminates noise information to achieve appealing noise reduction for industrial CT projections. This algorithm has relatively strong practicality due to an exemption from references and fewer computations compared to other iterative algorithms. And with the characteristics of "plug and play", it can be flexibly combined with other denoising algorithms to enhance the quality of CT reconstructed images further. As mentioned above, the feature value extraction is related to the value of the eigenvalue and the number of groupings; we just elaborate on the screening method relying on the value of the eigenvalue of the projection matrix. The grouping value $t$ in this paper is set as 4, which can be adjusted according to the number of samples.

\printbibliography

\end{document}